\documentclass[aps,prb,twocolumn,groupedaddress,showpacs]{revtex4}
\usepackage[dvips]{graphics}
\begin{document}
\title{Density Functional Study of Cubic to Rhombohedral Transition
       in $\alpha$-AlF$_3$}
\author{Yiing-Rei Chen$^1$, Vasili Perebeinos$^2$, and Philip B. Allen$^{1,3}$}
\affiliation{$^1$Department of Physics and Astronomy, State University
       of New York, Stony Brook, New York 11794-3800\\
       $^2$IBM Research Division, Thomas J. Watson Research Center, Yorktown 
       Heights, New York 10598 \\
       $^3$Department of Applied Physics and Applied Mathematics, and the
       Materials Research Science and Engineering Center, Columbia University,
       New York, New York 10027}
\date{\today}
\begin{abstract}
 Under heating, $\alpha$-AlF$_3$ undergoes a structural phase transition
 from rhombohedral to cubic at temperature $T$ around 730 K.  The density
 functional method is used to examine the $T$=0 energy surface in the
 structural parameter space, and finds the minimum in good agreement with
 the observed rhombohedral structure.  The energy surface and electronic
 wave-functions at the minimum are then used to calculate properties
 including density of states, $\Gamma$-point phonon modes, and the dielectric
 function.  The dipole formed at each fluorine ion in the low temperature
 phase is also calculated, and is used in a classical electrostatic picture
 to examine possible antiferroelectric aspects of this phase transition.
\end{abstract}  
\maketitle
\section{Introduction}
 
 The ionic insulator AlF$_3$ has a number of known polymorphs 
 \cite{phases1,phases2}, which all convert irreversibly to the stable 
 $\alpha$-AlF$_3$ within the temperature range approximately 730 to 920 K.  
 Recent interest in this material arises due to its catalytic activity for 
 dismutation and halogen exchange reactions \cite{catalytic}.

 Above its transition temperature (about 730 K)\cite{Grey}, $\alpha$-AlF$_3$
 has the cubic perovskite structure AMX$_3$, with the A cations absent (or
 the ReO$_3$ structure.)  Aluminum plays the role of the M cation, and is 
 surrounded by an octahedron of corner-shared fluorine atoms.  At low 
 temperature, the structure becomes rhombohedral, and this symmetry lowering 
 can be characterized as a rotation of the fluorine octahedron about one of 
 the three-fold axis of the perovskite cubic cell (or the $a^-a^-a^-$ system 
 in Glazer's tilt system \cite{Woodward}).  As every adjacent octahedron 
 rotates in the opposite sense, the wave-vector of this distortion is 
 ($\pi,\pi,\pi$), and the unit cell becomes double the size of that in the 
 cubic phase.

 We study the $\alpha$-AlF$_3$ structure by using pseudopotentials, a plane-wave
 basis, and the LDA method.  After being tested, the pseudopotentials
 are applied in bulk $\alpha$-AlF$_3$.  The total energy surface is examined
 in the structural parameter space where the cubic phase is compared with
 the rhombohedral phase found at the deepest nearby minimum of the energy
 surface.

\section{Testing Different Pseudopotentials}
 The Troullier and Martins method \cite{T&M} is used to generate the
 pseudopotentials.  Since there are different possibilities of choosing
 cut-off radii $r_{cl}$, as well as how many valence orbitals to include
 in the pseudopotentials,  we test the pseudopotentials by doing LDA
 self-consistent iterations for a crystal or molecule prototype and
 compare the results with FP-LAPW \cite{Blaha} all-electron calculations 
 (full potential linearized augmented plane-wave \cite{Singh}).  This 
 provides guidance on the choice of energy cut-off of the plane-wave basis, 
 and also the choice of local potential, in the $\alpha$-AlF$_3$ calculation
 that follows.  The results of these tests are shown in Table \ref{tab:Al} 
 and Table \ref{tab:F}, for aluminum and fluorine separately.  For the
 Ceperley-Alder exchange-correlation energy \cite{CA}, the Vosko-Wilk-Nusair
 \cite{VWN} parametrization is used.

 The test of the aluminum pseudopotential is performed in bulk Al metal, not
 only the observed {\it fcc} structure, but also the {\it sc} and
 {\it bcc} structures, where comparison can still be made between the
 all-electron method and LDA.  The results show that the pseudopotential
 for the Al$^{3+}$ ion should be capable of describing the unoccupied
 $3d$ orbital.  Otherwise, even the large $r_{cl}$ ($l=0,1$ for $s$ and
 $p$ orbitals) needed to fix the lattice constant in the {\it fcc} aluminum
 are insufficient to remove the $r_{cl}$-dependence of the lattice
 constant and bulk modulus.  Due to the more attractive nature of the $d$
 pseudopotential, including the $3d$ orbital reduces the dependence of
 these two quantities on $r_{cl}$ and the choice of local potential, and
 improves the agreement with the all-electron results.  Better agreement 
 is also seen in comparisons for $sc$ and {\it bcc} structures of Al metal.

 The fluorine pseudopotential is tested for the F$_2$ molecule.  Unlike
 the low plane-wave cut-off needed for aluminum (less than 40 Rydberg),
 more than 90 Rydberg is needed in the F$_2$ molecule.  For bulk Al metal 
 we use a k-point mesh of 12$\times$12$\times$12 that contains 56 special 
 k-points, while for F$_2$ we do a one k-point calculation.

 Results in the following sections are calculated using pseudopotentials
 (Al$^\dag$, F$^\dag$) as shown in the captions of the Tables.
 This set of pseudopotentials gives the lattice constant ($a = 5.02 \AA$),
 and bulk modulus ($B = 146.6$ GPa) of cubic $\alpha$-AlF$_3$ in
 good agreement with all-electron results ($a=5.043\AA$, $B=150$ GPa.)

\begin{table}
 \caption{Aluminum pseudopotential tests on bulk Al.  The first column
 shows the cut-off radii in units of Bohr radius ($a_0$) for the valence
 orbitals (l=0,1,2).  Tests with r$_{\rm c2}$ omitted do not include the
 $d$ orbital.  The following columns show the use of core correction, the
 choice of local potential, lattice constant, bulk modulus and the total
 energy per atom with respect to the {\it fcc} structure.  Values are shown
 in atomic units ($e=$ electron charge, $a_0=$ Bohr radius.)  $\Delta$E
 is defined to be zero for the {\it fcc} structure, while the total energy
 differences between {\it fcc} and other structures can be compared with
 all-electron results. The pseudopotential with a $^\dag$ is chosen for the
 calculations of $\alpha$-AlF$_3$. \label{tab:Al} }
 \begin{ruledtabular}
 \begin{tabular}{cccccc}
 \makebox[1.2cm]{r$_{\rm c0}$/r$_{\rm c1}$/r$_{\rm c2}$} &
 \makebox[0.8cm]{core} &
 \makebox[0.7cm]{local} &
 \makebox[1cm]{lattice} &
 \makebox[1cm]{B} &
 \makebox[1cm]{$\Delta$E} \\
 ($a_0$) & corr. && const.($a_0$) & ($\times$10$^{-3}e^2/a_0^4$) &
 ($e^2/a_0$)\\
 \hline
 {\it fcc} structure &&&& \\
 2.2/2.5/---  & yes & $s$ & 7.738 & 2.487 & 0 \\
 2.4/2.5/---  & yes & $s$ & 7.712 & 2.524 & 0 \\
 2.6/2.6/---  & yes & $s$ & 7.672 & 2.607 & 0 \\
 1.9/2.4/2.7  & yes & $d$ & 7.507 & 2.873 & 0 \\
 1.8/2.2/2.7  & yes & $s$ & 7.500 & 2.850 & 0 \\
 1.8/2.2/2.7  & yes & $p$ & 7.516 & 2.896 & 0 \\
 1.8/2.2/2.7 & yes & $d$ & 7.507 & 2.873 & 0 \\
 1.8/2.2/2.6  & yes & $d$ & 7.505 & 2.934 & 0 \\
 2.6/2.6/2.6$^\dag$ & no & $d$ & 7.481 & 2.865 & 0 \\
 all-electron & --- & --- & 7.536 & 2.905 & 0 \\
 experiment   & --- & --- & 7.650  & 2.579 & --- \\
 \hline
 {\it sc} structure &&&& \\
 1.9/2.4/2.7  & yes & $d$ & 5.063 & 2.176 & 0.01431 \\
 1.8/2.2/2.7  & yes & $s$ & 5.061 & 2.170 & 0.01451 \\
 1.8/2.2/2.7  & yes & $p$ & 5.069 & 2.195 & 0.01425 \\
 1.8/2.2/2.7  & yes & $d$ & 5.063 & 2.177 & 0.01431 \\
 1.8/2.2/2.6  & yes & $d$ & 5.062 & 2.175 & 0.01436 \\
 2.6/2.6/2.6$^\dag$   & no & $d$ & 5.050 &2.072 & 0.01462 \\
 all-electron & --- & --- & 5.065 & 2.096 & 0.01449 \\
 \hline
 {\it bcc} structure &&&& \\
 1.9/2.4/2.7  & yes & $d$ & 6.015 & 2.546 & 0.00368 \\
 1.8/2.2/2.7  & yes & $s$ & 6.009 & 2.534 & 0.00374 \\
 1.8/2.2/2.7  & yes & $p$ & 6.023 & 2.581 & 0.00368 \\
 1.8/2.2/2.7  & yes & $d$ & 6.015 & 2.542 & 0.00368 \\
 1.8/2.2/2.6  & yes & $d$ & 6.014 & 2.538 & 0.00370 \\
 2.6/2.6/2.6$^\dag$   & no & $d$ & 5.992 & 2.472 & 0.00380 \\
 all-electron & --- & --- & 6.027 & 2.525 & 0.00414 \\
 \end{tabular}
 \end{ruledtabular}
\end{table}
\begin{table}
 \caption{Fluorine pseudopotential tests on F$_2$ molecule.  The last
 column indicates the plane-wave cut-off needed to obtain the convergent
 result.  The pseudopotential with a $^\dag$ is chosen for the calculations
 of $\alpha$-AlF$_3$.  \label{tab:F}}
 \begin{ruledtabular}
 \begin{tabular}{lccccc}
 \makebox[1.5cm]{r$_{\rm c0}$/r$_{\rm c1}$} &
 \makebox[0.8cm]{core} &
 \makebox[0.7cm]{local} &
 \makebox[1.0cm]{bond} &
 \makebox[1.0cm]{force} &
 \makebox[1.0cm]{cut-off} \\
 ($a_0$) & corr. && ($a_0$) & const.($e^2/a_0^3$) & (Ryd.) \\
 \hline
 1.4/1.6      & no  & $p$ & 2.780 & 0.370 & 90  \\
 1.4/1.45     & yes & $p$ & 2.710 & 0.408 & 90  \\
 1.3/1.3      & yes & $s$ & 2.599 & 0.384 & over 100  \\
 1.3/1.3$^\dag$  & no  & $s$ & 2.603 & 0.376 & 90  \\
 1.2/1.2      & yes & $s$ & 2.617 & 0.387 & over 100 \\
 all-electron & --- & --- & 2.632 & 0.376 & --- \\
 experiment   & --- & --- & 2.669 & 0.302 & --- \\
 \end{tabular}
 \end{ruledtabular}
\end{table}
\section{Results from LDA}
\subsection{k-point Test and Density of States}
\begin{figure}[ht]
 \centerline{\scalebox{0.33}[0.33]{\includegraphics*{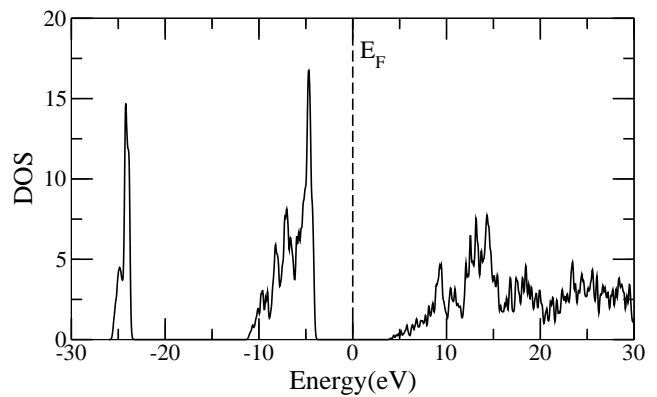}}}
 \caption{Density of states plot of $\alpha$-AlF$_3$ in the rhombohedral
 ground state, using pseudopotential (Al$^\dag$, F$^\dag$). The
 rhombohedral ground state has unit cell edge 4.76$\AA$,
 $\theta$=57.5$^\circ$ and $\delta$=0.098.  This calculation
 includes 50 states and uses a k-point mesh 6$\times$6$\times$6.
 The temperature broadening is 54 meV. \label{fig:dos}}
\end{figure}

 A k-point mesh of 3$\times$3$\times$3 is used for the Brillouin zone 
 summation for $\alpha$-AlF$_3$.  This is tested for cubic $\alpha$-AlF$_3$,
 where a finer mesh of 4$\times$4$\times$4 gives a barely different energy
 {\it vs} plane wave cut-off curve, with energy uncertainty less than 1 meV. 
 This mesh gives six special k-points after symmetrization (according to the
 point group D$_{\rm 3d}$ of the crystal), over which we include at least
 36 states in the truncated Hamiltonian diagonalization processes.  Since in
 the rhombohedral unit cell of $\alpha$-AlF$_3$ we consider the 48 valence
 electrons from aluminum $3s^{2}3p^{1}$ and fluorine $2s^{2}2p^{5}$, these
 electrons will occupy the 24 lowest lying bands.  A test including more
 states, 48 states for example, shows that in the process of iterative
 diagonalization, using 36 states gives convergent answers for the eigenvalues
 of the lowest 24 bands.  Thus we include 36 states in the calculation of the 
 minimum on the energy surface.  To make the density of states plot, 
 we include 50 states to better describe the empty states from aluminum 
 orbitals.  As shown in Fig. \ref{fig:dos},  LDA gives an insulator 
 band-gap of 8 eV.  We choose 90 Rydberg to be the energy cut-off for the 
 plane-wave basis.  This choice gives the same energy difference between 
 the two structures as that from the choice of 140 Rydberg (with error 
 within $1\sim 2$ meV.  The two structures are the cubic and the rhombohedral 
 structure at the minimum.)
\subsection{Crystal Structure}
 We use LDA to search for the ground state in the rhombohedral symmetry
 class $R\overline{3}C$ seen experimentally in $\alpha$-AlF$_3$\cite{Grey}.  
 There are three free parameters: (1) the lattice constant $a$, (2) the angle 
 $\theta$ between any two of the rhombohedral lattice vectors, and (3) the 
 inner parameter $\delta$, which describes the bending of the Al-F-Al bond.  
 To describe these three parameters, we first start from the cubic structure 
 with lattice vectors $(2b,0,0)$, $(0,2b,0)$ and $(0,0,2b)$, where $b$ is the 
 Al-F bond length.  A rhombohedral unit cell is formed by picking new lattice 
 vectors $\vec{a}_1=(0,2b,2b)$, $\vec{a}_2=(2b,0,2b)$, and $\vec{a}_3=(2b,2b,0)$.  
 The angle $\theta$ defined as 
 $\theta = \cos^{-1}(\vec{a}_1\cdot\vec{a}_2/|\vec{a}_1||\vec{a}_2|)$
 is $60^\circ$ in this case.  This new cell is then elongated along the [111] 
 direction.  Consequently, the strains 
 $\epsilon_{23}=\epsilon_{31}=\epsilon_{12}$ become nonzero, and 
 $\theta\neq 60^\circ$.  Viewed from the [111] direction, in the aluminum plane  
 the ions form a hexagonal pattern, and the nearest Al-Al distance on this 
 (111) plane is the lattice constant $a$.  This rhombohedral structure can also
 be described in a hexagonal setting, with the following $c/a$ ratio
\begin{equation}
\frac{c}{a} = \sqrt{\left(\frac32\right)
\left(\frac{1+2\cos\theta}{1-\cos\theta}\right)}
\end{equation}
 The cell shape and cell volume $V$ are determined by $a$ and $c/a$:
\begin{equation}
 V = \frac{a^3}{2\sqrt{2}}\sqrt{\frac{1+2\cos\theta}{1-\cos\theta}} 
   = \frac{a^2c}{2\sqrt{3}},
\end{equation}
 while the relation between the lattice constant $a$ and the length of
 the rhombohedral lattice vectors is:
\begin{equation}
 a / |\vec{a}_1| = \sqrt{2(1-\cos\theta)}
\end{equation}
 There are two aluminum ions and six fluorine ions in each rhombohedral
 unit cell.  The fluorine ions sit at the 6e sites ($(x,\bar{x}+\frac 12,
 \frac 14)$, $(x,x+\frac 12,\frac 34)$, etc.), and $\delta$=$x$-0.75 is  
 the deviation from $x$=0.75 where the Al-F-Al bond angle is strictly    
 180$^\circ$.  The distortion caused by $\delta$ alone defines the octahedron 
 rotation angle $\omega$ about the [111] axis:
\begin{equation}
 \tan\omega = 2\sqrt{3}\delta,
\label{eq:delta}
\end{equation}
 On the other hand, the $c/a$ ratio and the $\omega$ have the following 
 relation:
\begin{equation}
 \frac{c}{\sqrt{6}a} = \frac{1+\zeta}{\cos\omega},
\label{eq:zeta} 
\end{equation}
 where $\zeta$ describes the flattening ($\zeta<0$) and elongation ($\zeta>0$)
 of the octahedron along [111].  

\begin{figure}[htbp]
 \centerline{\scalebox{0.32}[0.32]{\includegraphics*{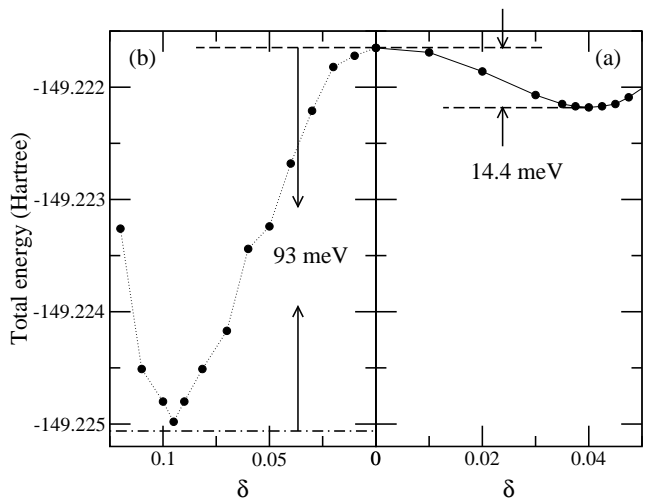}}}
 \caption{The relaxation of $\delta$ from the cubic structure is shown
 in plot (a).  The curve of total energy (per two formulas of AlF$_3$)
 {\it vs} $\delta$ at $\theta=60.0^\circ$ and $a=5.02\AA$ shows that 
 the cubic phase is sitting at an energy maximum.  In (b), the curve
 follows the rigid-octahedron path, and leads to the region near the
 overall minimum in the parameter space.  The dotted-dashed line shows the
 energy of the overall minimum.}
 \label{fig:delta}
\end{figure}

 We first assume cubic symmetry by fixing $\theta=60^\circ$  
 and $\delta=0$.  LDA gives the lattice constant $a=a^*=5.02\AA$ for 
 the cubic phase.  With the lattice constant kept fixed at the value $a = a^*$,
 $\delta$ is then relaxed to give minimum energy
 at $(a, \theta, \delta) = (a^*, 60^\circ, \pm 0.04)$.  As shown in
 Fig. \ref{fig:delta}, the relaxation of $\delta$ finds minima at
 $\delta$=$\pm0.04$ and a maximum at $\delta=0$, with an energy difference
 of 14.4 meV, showing that in our $T$=0 energy surface study, the unbent
 Al-F-Al bond, and therefore the strictly cubic structure, is not   
 a metastable solution.

 The three parameters are then relaxed in turn to approach the deepest
 nearby minimum.  In fact, in the parameter space, there is a path that keeps
 the fluorine octahedra rigid and leads to the region near the overall 
 minimum.  A Rigid octahedron means that the Al-F bonds are fixed at the bond 
 length found by LDA in the cubic phase
\begin{equation}
 b = \frac{a^*}{2\sqrt{2}} = \frac{a}{2\sqrt{2} \cos\omega}
\end{equation}
 and that the angles between neighboring Al-F bonds are always 90 degrees 
 ($\zeta=0$).  With these two restrictions, only one parameter $\delta$ 
 remains and defines the rigid octahedron path.  The energy along this 
 path is also shown in Fig. \ref{fig:delta}.  The overall minimum locates 
 not far from the minimum on this path.  It has an Al-F bond length only 
 $0.3\%$ longer, a pretty small octahedron elongation factor $\zeta = 0.25\%$, 
 and the total energy 93 meV lower than the cubic phase.  
 Table \ref{tab:min} shows the comparisons between LDA and experimental data 
 \cite{Grey}.  While no $T=0$ experimental numbers are available, 
 one can only make reasonable comparisons for the deviation of bond length and 
 $\zeta$.  The rhombohdral phase observed in the experiment has an Al-F bond 
 length $0.43\%$ longer than the value observed in the cubic phase, and 
 $\zeta = -0.102\%$ (meaning the octahedron is slightly flattened.)  This 
 indicates that both the LDA and the experimentally found rhombohedral phases 
 are close to the rigid  octahedron path.
 \begin{table}[here]
 \caption{The structure parameters found by LDA are compared with
  experimental values \cite{Grey} here for different phases.  The total 
  energy (per two formulas of AlF$_3$) difference between the two phases 
  is also shown here, where the total energy of the cubic phase is set
  to zero. The values for the rhombohedral phase are calculated at
  the minimum of the energy surface.  \label{tab:min}}
 \begin{ruledtabular}
 \begin{tabular}{lcccc}
 \makebox[2.2cm] &
 \makebox[0.8cm]{lattice} &
 \makebox[0.4cm]{$\theta(^\circ$)} &
 \makebox[0.4cm]{$\delta$} &
 \makebox[0.4cm]{$\Delta$E} \\
 & const.($\AA$)&&&(meV) \\
 \hline
 psp(Al$^\dag$, F$^\dag$) rhom  & 4.76   & 57.5  & 0.098 & -93  \\
 experiment (LT)                & 4.9382 & 58.82 & 0.0691 & ---   \\
 psp(Al$^\dag$, F$^\dag$) cubic & 5.02   & 60.0  & 0.0 & --- \\
 experiment (HT)                & 5.0549 & 59.94 & 0.0129 & ---   \\
 \end{tabular}
 \end{ruledtabular}
\end{table}
\subsection{Phonon Modes}
 Previous workers \cite{MF3} investigated the phonon spectrum of the 
 cubic phase using the generalized Gordon-Kim method\cite{Ivanov}.
 Density functional theory was applied to construct the charge
 distribution and polarizability of ion.  An approximate crystal energy 
 is then calculated from the ion results, and is used to examine
 the crystal lattice dynamics.  However they did not find an instability 
 of the cubic phase, which is contradictory to the experiment.  Here we use 
 full LDA to examine the energy surface around the rhombohedral minimum, 
 and extract the soft phonon modes directly.  The rhombohedral minimum 
 described in the previous section, and the dipole aspect discussed 
 in Sec. \ref{sec:dipole}, both show the instability of the cubic phase. 

 At the $\Gamma$ point of the rhombohedral phase, the irreducible 
 representations are those of the point group $D_{3d}$ (note that the 
 space group of the rhombohedral phase is $R\overline{3}C$, which is 
 nonsymmorphic.)  Among the 12 $\Gamma$ point modes provided by the fluorines,
 namely, $A_{1{\rm g}} \oplus 2A_{2{\rm g}} \oplus 3E_{\rm g} \oplus A_{1\rm u}
 \oplus 2A_{\rm u} \oplus 3E_{\rm u}$, the four modes 
 $A_{1{\rm g}} \oplus 3E_{\rm g}$ are Raman active.  Viewed from the cubic 
 phase, the octahedron rotation that governs the structural transition is the 
 R$_5$ mode (in Kovalev labelling) at the zone boundary 
 ($\pi,\pi,\pi$)\cite{Raman1}.  After distortion, the R$_5$ mode gives rise 
 to the zone-center A$_{\rm 1g}$ mode and one of the three E$_{\rm g}$ modes.  
 Raman experiments \cite{Raman1, Raman2} showed that the A$_{\rm 1g}$ and one 
 E$_{\rm g}$ are the soft phonon modes below transition.  The A$_{\rm 1g}$ is 
 the Al-F-Al bond angle bending mode.  Using the fluorine atomic mass, and the 
 energy {\it vs} $\delta$ curve, the A$_{\rm 1g}$ frequency is calculated to be 
 205 cm$^{-1}$.  We also calculate the other restoring coefficients of the 
 three E$_{g}$ modes, and the two Raman-inactive A$_{\rm 2g}$ 
 modes.  Their frequencies are listed in Table. \ref{tab:freq}.
\begin{table} 
 \caption{Some of the $\Gamma$ point phonon modes frequencies calculated by LDA.  
  direct comparison with experiments are not available since previous
  observations are done at RT and above.  However, in the
  experiment it is found that the two high energy E$_{\rm g}$ modes only 
  weakly depend on temperature.  A crude extrapolation using 
  $\omega^2 = A(T-T_c)$ (from Curie-Weiss law and Lyddane-Sachs-Teller relation)
  and the RT experimental data\cite{Raman1} gives the frequency values at 
  $T=0$. \label{tab:freq}}
 \begin{ruledtabular}
 \begin{tabular}{lcccc}
 \makebox[1.0cm] &
 \makebox[0.6cm]{A$_{1g}$}  &
 \makebox[0.6cm]{E$_g$} &
 \makebox[0.6cm]{E$_g$} &
 \makebox[0.6cm]{E$_g$} \\
 \hline \\
 LDA            & 205 & 182 & 350 & 487 \\
 experiment(RT) & 158 &  98 & 383 & 481 \\
 extrapolation  & 190 & 118 & 388 & 481 \\
 \hline\hline
 & A$_{2g}$ & A$_{2g}$ & & \\
 \hline
 LDA            & 361 & 691 & & \\
 \end{tabular}
 \end{ruledtabular}
\end{table}

\subsection{Dielectric Function}
\begin{figure}[htbp]
 \centerline{\scalebox{0.35}[0.35]{\includegraphics*{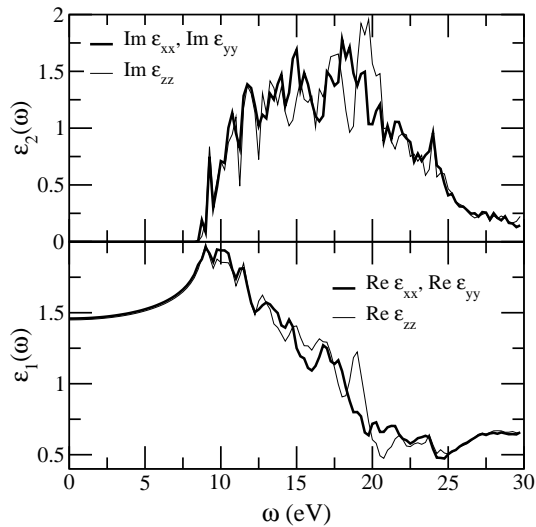}}}
 \caption{Plots of dielectric functions.  This calculation includes 
  50 states and uses a k-point mesh $6\times6\times6$.}
 \label{fig:dielectric}
\end{figure}

 The electronic part of the dielectric tensor
 $\epsilon(\mbox{\boldmath $q$}=0,\omega)$ is also calculated.
 $\epsilon_{xx} = \epsilon_{yy} \not= \epsilon_{zz}$ is expected since
 we have chosen lattice vectors such that $\hat x$ and $\hat y$ are
 perpendicular to the three-fold axis.  The imaginary part of the dielectric
 function can be directly obtained by calculating all direct inter-band
 transitions \cite{Yu}:
\begin{eqnarray}
 \epsilon_2(\omega) &=& (\frac{2\pi e}{m\omega})^2
 \sum_{k}\sum_{m,n}\left|\langle\psi_{n,k}\left|\vec p
 \right|\psi_{m,k}\rangle\right|^2
 f_n(1-f_m) \nonumber \\
 &\times& \delta(E_m-E_n-\hbar\omega),
 \end{eqnarray}
 where $\vec p$ is the momentum operator, and k runs through all k-points 
 in the Brillouin zone allowed in a unit volume.  The band indices are 
 $m$ and $n$, $f_{n,k}$ and $f_{m,k}$ are the occupation number of the
 $n^{\rm th}$ and $m^{\rm th}$ states at the $k^{\rm th}$ k-point.  The real
 part of dielectric function is obtained using the Kramers-Kronig relation.  
 The result at $T$=0 is given in Fig. \ref{fig:dielectric}, but we do not know
 of any experiment to compare with.  However this calculation does neglect 
 the non-local potential effect \cite{nonlocalV} and the local field effect
 \cite{localE}, and follows the pseudo-wave-functions that are smoothened 
 in the ion core regions.
\section{Inner Parameter $\delta$ and Dipole Formation}
\label{sec:dipole}
 The formation of a dipole plays an important role in the energy difference
 between the two $\alpha$-AlF$_3$ structures.  When the inner parameter
 $\delta$ is non-zero, each fluorine atom develops a dipole (Aluminum
 atoms sit on inversion centers and therefore cannot have dipoles).
 Starting from the cubic structure, the distortion from $\delta$ alone
 defines the octahedron rotation angle $\omega$ about the [111] axis 
 (see Eq. \ref{eq:delta}) and already lowers the symmetry to $R\overline{3}C$.  
 The parameter $\delta$ is irrelevant to the $c/a$ ratio, but it does flatten 
 the octahedron (see Eq. \ref{eq:zeta}).  It also elongates the Al-F bond 
 length, and introduces a polarization to each fluorine.  The energy gain from 
 the fluorine dipole-dipole interaction drives the structure off-cubic.  This 
 is shown in Fig.\ref{fig:delta}, where the cubic structure is found to be not 
 a metastable solution.

\begin{figure}[bpt]
 \centerline{\scalebox{0.42}[0.42]{\includegraphics*{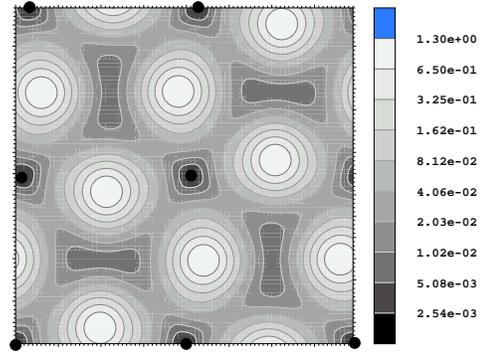}}}
 \caption{Charge density plot of the rhombohedral phase on a (100) plane.  
 The [100] direction is ill-defined since it is inherited from the cubic 
 phase, but the plane shown here does contain the aluminum ions (labelled 
 in solid dots) that form the (100) plane in the cubic phase.  The
 fluorine ions are slightly off the plane shown here, and display clearly
 a prolate charge distribution.
 \label{fig:CDP}}
\end{figure}

 A naive picture of ionic solids has spherical electron charge clouds of 
 total charge Q$_i$ around the $i^{\rm th}$ ion.  In reality, charge clouds
 are distorted.  For example, in high T cubic $\alpha$-AlF$_3$, the F ions
 are noticeably prolate when examined in a (100)-plane charge contour 
 calculation \cite{Madden}.  This distortion is still evident in the low T
 rhombohedral phase, as shown in Fig. \ref{fig:CDP}.  The value of the charge 
 Q$_i$, on the other hand, is not uniquely definable.  By creating a sphere 
 centered at each fluorine atom, with a radius that only allows neighboring
 spheres in the rhombohedral phase to touch each other at one point, we define
 a volume to examine the charge and the dipole moment of the fluorine.
 In the cubic phase, such a sphere contains charge -0.675$|e|$, while it
 contains -0.68$|e|$ in the rhombohedral phase (with a slightly different 
 radius.)  Using the same sphere, we compute the local induced dipole moment 
 $\mbox{{\boldmath ${p}$}}_{\rm ind}$ of the fluorine ion (which also has no 
 unique definition \cite{AlCl3}.)  As expected from symmetry, $p_{\rm ind}=0$ 
 in the cubic phase, whereas $p_{\rm ind}=0.103 |e|\AA$ in the rhombohedral 
 phase, pointing from the fluorine position of the unbent Al-F-Al bond towards 
 the actual distorted fluorine position.  Note that the displacive dipole 
 is $p_{\rm dis} = 0.315 |e|\AA$, defined by the fluorine charge -0.675$|e|$ 
 and the displacement $0.467\AA$ of the fluorine ion when the Al-F-Al bonds 
 bend.  By symmetry, the other fluorine sitting at the opposite side of the 
 octahedron has its dipole pointing exactly in the opposite direction.  
 Therefore the material could be called antiferroelectric \cite{AFE}.

 The dipoles and charges can as well be determined by doing integration in
 one of the equivalent polyhedra that surround the F$^-$ ions and partition 
 the whole space (the Al$^{3+}$'s sit on shared vertices of the F$^-$ 
 polyhedra.)  In this way, the ions are charged as Al$^{3+}$ and F$^-$,
 while $p_{\rm ind}=0.0963 |e|\AA$ and $p_{\rm dis} = 0.467 |e|\AA$.

 It is worthwhile to see if a simplified picture with these numbers
 helps to understand the energy.  In LDA, the energy includes
 the exact electrostatic or Hartree energy from the electron charge
 clouds, as well as other quantum effects (energy of delocalization,
 exchange and correlation.)  Purely classical electrostatic models (plus
 hard core repulsion), on the other hand, provide a simple but useful
 view.  We shall see how much classical electrostatic energy comes
 from the charge and if this helps to explain the stability of the
 structural deformation.

 In a simplified picture where only electrostatic energy and dipole
 formation energy ($E_1$ in Eq. \ref{eq:madel}, where $\alpha = 0.858 \AA^3$ 
 \cite{Kittel}) are considered, the total energy is
\begin{eqnarray}
 E &=& E_1 + E_2 + E_3 + E_4 \nonumber \\
 &=& \sum_i\frac{\mbox{\boldmath $p$}^2_i}{2\alpha}
  + \sum_i\sum_{j<i}\frac{Z_iZ_j}{r_{ij}}
  + \sum_i\sum_{j<i}\frac{Z_i(\mbox{\boldmath $p$}_j
    \cdot \mbox{\boldmath $r$}_{ij})}{r^3_{ij}}
 \nonumber \\
 &+& \sum_i\sum_{j<i}\frac{\mbox{\boldmath $p$}_i
 \cdot \mbox{\boldmath $p$}_j-3(\mbox{\boldmath $\hat r$}_{ij}
 \cdot \mbox{\boldmath $p$}_i)(\mbox{\boldmath $\hat r$}_{ij}
 \cdot \mbox{\boldmath $p$}_j)}{r^3_{ij}}.
 \label{eq:madel}
\end{eqnarray}
 This form leaves out the quantum effects and the effect from covalent 
 bond angle.  We find that $E_2$, which is purely ionic electrostatic 
 (including displacive dipoles), is less negative in the rhombohedral
 structure than in the cubic structure (i.e., if the same ion charge
 assignment is considered in both cases.)  However, when the 
 induced-dipole-related terms ($E_1+E_3+E_4$) are also considered, the energy
 is lowered.  By assuming point charges and point dipoles, we use the results
 from the polyhedron method to illustrate this picture: $E_2$ per two AlF$_3$
 formulas in the rhombohedral phase is 1 eV higher than that in the cubic 
 phase; but $E_3$ and $E_4$ bring the energy $E$ down, with the cost of $E_1$,
 to 0.97 eV lower than the cubic phase (which has only $E_2$.) 
 This suggests that the effect of induced dipoles is more than enough to 
 compensate the energy loss from the structural distortion.
%

\section{Conclusion}
 We report an LDA study of bulk $\alpha$-AlF$_3$.  By examining the $T$=0 energy 
 surface for structures of the phases on both sides of the transition, we find 
 the structural parameters to agree with previous experiments, and the cubic
 phase not to be a metastable solution.  Using the result of LDA, the density of
 states plot and the dielectric function are provided.  At the $\Gamma$-point, 
 the predicted A$_{\rm 1g}$ soft phonon mode and E$_{\rm g}$ modes are compared 
 with previous Raman experiment, while two other A$_{\rm 2g}$ modes are predicted.  
 We look at the charge and dipole moment at each fluorine ion, and use these 
 quantities to calculate the classical electrostatic energy.  In the 
 antiferroelectric distortion which accompanies the structural transition, 
 a classical calculation shows that the electrostatic energy gain from dipoles 
 is more than enough to compensates the energy loss from the ion-array deformation.

\acknowledgments
 We thank C. Grey and S. Chaudhuri for suggesting the project and for
 helpful discussions.  The LDA code (BEST) and the FT-LAPW code is provided 
 by Brookhaven National Laboratory.  Work at Stony Brook was supported in 
 part by NSF grant no. DMR-0089492.  Work at Columbia was supported in part 
 by the MRSEC Program of the National Science Foundation under Award Number 
 DMR-0213574.


\begin{references}
\bibitem{phases1}  N. Herron, D. L. Thorn, R. L. Harlow,
                   G. A. Jones, J. B. Parise, J. A. Fernandez-Baca,
                   and T. Vogt, Chem. Mater. {\bf 7}, 75 (1995).
\bibitem{phases2}  C. Alonso, A. Morato, F. Medina, F. Guirado, Y. Cesteros,
                   P. Salagre, and J. E. Sueiras, Chem. Mater. {\bf 12},
                   1148 (2000).
\bibitem{catalytic} E. Kemnitz and L. E. Manzer, Prog. Solid State Chem., 
                   {\bf 26} 97 (1998).
\bibitem{Grey}     P. J. Chupas, M. F. Ciraolo, J. C. Hanson, and C. P. Grey,
                   J. Am. Chem. Soc. {\bf 123}, 1694 (2001).
\bibitem{Woodward} P. M. Woodward, Acta Cryst. {\bf B53}, 32 (1997).
\bibitem{T&M}      N. Troullier and J. L. Martins,
                   Phys. Rev. B {\bf 43}, 1993 (1991).
\bibitem{Blaha}    P. Blaha, K. Schwarz, and J. Luitz, 
                   in {\it Proceedings of WIEN97}
                   (Techn. Universit\"{a}t Wien, Austria, 1999).
\bibitem{Singh}    D. J. Singh, {\it Planewaves, Pseudopotentials
                   and the LAPW Method} (Kluwer Academic, Boston, 1994).
\bibitem{CA}       D. M. Ceperley and B. J. Alder.
                   Phys. Rev. Lett., {\bf 45}, 566, 1980.
\bibitem{VWN}      S. H. Vosko, L. Wilk, and M. Nusair, 
                   Can. J. Phys., {\bf 58}, 1200, 1980.
                   L.~Wilk and S.~H. Vosko, 
                   J. Phys. C Solid State, {\bf 15}, 2139, 1982. 
\bibitem{MF3}      V. I. Zinenko and M. G. Zamkova, Phys. Solid State
                   {\bf 42}, 1348 (2000).
\bibitem{Ivanov}   O. V. Ivanov and E. G. Maksimov, JETP {\bf 81} 1008 (1995).
\bibitem{Raman1}   P. Danial, A. Bulou, M. Rousseau, J. Nouet, J. L. Fourquet,
                   M. Leblanc, and R. Burriel, J. Phys.: Condens. Matter
                   {\bf 2}, 5663 (1990).
\bibitem{Raman2}   P. Danial, A. Bulou, M. Rousseau, and J. Nouet, Phys. Rev.
                   B {\bf 42}, 10545 (1990).
\bibitem{Yu}       P. Yu and M. Cardona, {\it Fundamentals of Semiconductors},
                   second edition, page 251. (Springer, 1999)
\bibitem{nonlocalV}  B. Adolph, V. I. Gavrilenko, K. Tenelson, F. Bechstedt
                   and R. Del Sole, Phys. Rev. B {\bf 53} 9797 (1996).
\bibitem{localE}   M. S. Hybertsen and S. G. Louie, Phys. Rev. B {\bf 35}
                   5585 (1987).
\bibitem{Madden}   This was pointed out to us by P. Madden (private
                   communication), and later also appeared in our charge
                   contour plot.
\bibitem{AlCl3}    L. Bernasconi, P. A. Madden and M. Wilson,
                   Phys. Chem. Comm. {\bf 5}, 1 (2002).
\bibitem{AFE}      R. Blinc and B. \v{Z}e\v{k}s, {\it Soft Modes
                   in Ferroelectrics and Antiferroelectrics},
                   edited by E. P. Wohifarth 
                   (North-Holland Publishing Co. Amsterdam, 1974)
\bibitem{Kittel}   C. Kittel, {\it Introduction to Solid State Physics},
                   seventh edition, page 391 and references therein, John
                   Wiley and Son, New York.
\end{references}
\end{document}